\documentclass[prb,aps,amsmath,amssymb,preprintnumbers,twocolumn]{revtex4}
\usepackage{graphicx}
\usepackage{dcolumn}
\usepackage{bm}
 \begin{document}
 \preprint{{\it Applied Physics Letters in press}}
 \title{Performance of a spin-based insulated gate field effect transistor}
 \author{Kimberley C. Hall}
 \affiliation{Department of Physics, Dalhousie University, Dalhousie, Canada}
\author  {Michael E. Flatt\'e}
\affiliation {Optical Science and Technology Center and Department of Physics and Astronomy, University of Iowa, Iowa City,
IA 52242, USA}
 \begin{abstract}
Fundamental physical properties limiting the performance of spin field effect transistors are compared to those of ordinary (charge-based) field effect transistors. Instead of raising and lowering a barrier to current flow these spin transistors use static spin-selective barriers and gate control of spin relaxation. The different origins of transistor action lead to distinct size dependences of the power dissipation in these transistors and permit sufficiently small spin-based transistors to surpass the performance of charge-based transistors at room temperature or above. This includes lower threshold voltages, smaller gate capacitances, reduced gate switching energies and smaller source-drain leakage currents.
 \end{abstract}
 \maketitle

Spin-based electronic devices currently have broad commercial applications to magnetic field sensors and non-volatile memory devices\cite{Wolf:2001,Ziese}. Semiconductor spin-based electronic devices\cite{Spintronicsbook} have been shown to permit switching, modulation and gain, along with new functionality (principally non-volatility and spin-selective properties)\cite{Datta:1990,FV:2001,FYJA:2003,Schliemann:2003,Hall:2003}. As the management of active and leakage power dissipation is a key roadblock to scaling of traditional charge-based transistors beyond 2010\cite{class2004,narendra2004,ITRS:2003}, assertions\cite{Wolf:2001,Spintronicsbook} that spin-based devices may permit lower-power operation through the incorporation of reconfigurable logic chips into devices, or lower-power spin-based switching, have attracted considerable attention. Despite this, no quantitative comparisons of the key elements of transistor power dissipation, the leakage current and gate switching energies, have been performed between spin-based insulated gate field-effect transistors and charge-based metal oxide semiconductor field-effect transistors (MOSFETs) (although Ref.~\onlinecite{BC:2004} reports some narrowly-focused calculations). 

Here the performance of an individual spin transistor device is directly compared with current and future MOSFETs. This comparison relies on calculations of the leakage current and gate switching energy, in addition to the gate switching speed, source-drain saturation current and gate capacitance for a spin transistor. The semiconductor roadmap\cite{ITRS:2003} identifies three principal paths for complementary metal oxide semiconductor (CMOS) transistor structures: high-performance, low operating power, and low standby power designs. As the focus here is on fundamental power dissipation limits the comparisons here will consider those CMOS transistors with the most stringent power requirements, the low standby power (LSTP) development path. A principal conclusion is that the leakage current and switching energies of the spin transistor can be made significantly smaller than those of current {\it and future} LSTP CMOS transistors, including those scheduled for introduction on the semiconductor roadmap\cite{ITRS:2003} in 2018. This superior performance is tied to fundamental aspects of spin-based switching in an individual device. Some essential challenges that need to be overcome in order to achieve this level of performance in a spin transistor are also identified.

In order to make a direct comparison at the individual transistor level,  a spin transistor design is considered whose source, drain, and gate contacts are in local equilibrium. Thus the spin transistor cannot pass on a quantum-mechanically coherent current to the next transistor in a circuit, such as would be the case, {\it e.g.}, if the next transistor in the circuit used the spin polarization of the drain current of the previous transistor. This significantly restricts the potential designs for spintronic logic. Better performance might be achievable for a circuit using these more general designs than would be predicted based on individual transistor performance, however this approach still illuminates many key potential advantages of spin transistors.

\begin{figure}
\includegraphics[width=\columnwidth]{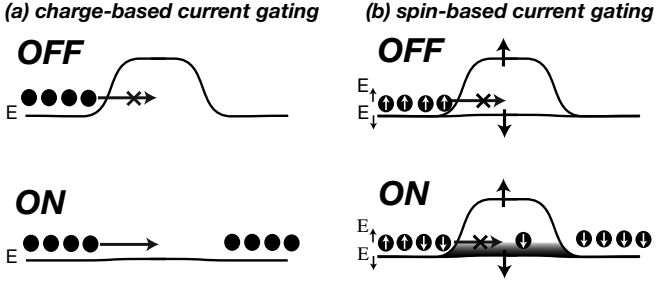}
\caption{Schematic barriers used in (a) a MOSFET and (b) a spin-based FET. A MOSFET works by controlling the height of the barrier, with a  barrier height and width largely determined by the desired on-off current ratio and leakage current.  The spin-based FET considered here works by controlling the nature of the initial state moving past the barrier in (b); if the initial state is fully spin polarized the transistor is off, otherwise it is on.}
\label{chargevspin}
\end{figure}

The role of the barrier to current flow differs qualitatively in the two FET designs. Shown in Fig.~\ref{chargevspin}(a) is a schematic of the ``off'' and the ``on'' positions of the barrier in a MOSFET. The electrons attempt to move from left to right (in a MOSFET this barrier is between the source and the drain) through a channel which is either insulating (off) or conducting (on). The height of the barrier, $V_{th}$, is controlled by a gate contact. For LSTP CMOS the barrier is designed to be at least 400~mV high, corresponding to $\sim 16 k_BT$ at room temperature, where $k_B$ is Boltzmann's constant and $T$ is the temperature. This is the minimum barrier height to reduce the ratio of the thermally excited current over the barrier in the off state to the current in the on state to $\sim 10^{-7}$.   Another central characteristic of the MOSFET is the gate capacitance $C_g$, which is proportional to the area of the region of the channel that is blocked with this barrier. The switching energy is  $C_gV_{th}^2/2$ (half the power-delay product\cite{ITRS:2003}) and the switching time is proportional to $C_g$. If the gate capacitance is too low the barrier becomes thin enough that carriers can tunnel through it and the leakage current rises, but if the capacitance is too high the switching time is long and the switching energy high.

The spin transistor design considered here is based on the spin-dependent barrier shown in Fig.~\ref{chargevspin}(b). A high, thick barrier is present for one spin orientation (shown as spin-up in the figure), and no barrier is present for the other spin orientation (shown as spin-down). Such a spin-dependent barrier may be realized, for example, using a half-metallic ferromagnetic contact\cite{HM} or a spin-selective resonant tunneling diode\cite{Voskoboynikov:2000,deAndradaeSilva:1999,Koga:PRL2002,Ting:2002,Hall:2003}. If the carriers attempting to move through the barrier are entirely polarized spin-up then they cannot move through the barrier. If the carriers are polarized spin-down, or are a mixture of spin-up and spin-down, then carriers can move through the barrier with ease. Switching the transistor from on to off consists of switching the carrier orientation from fully polarized spin-up to unpolarized with a gate field. As switching the transistor does not involve raising and lowering a barrier, the barrier for spin-up carriers can be much higher than 400~mV and can be thick, without negative consequences for the on-off ratio or the leakage current. CMOS's tradeoff between dynamic and static power dissipation, which represents a central roadblock to scaling\cite{class2004,narendra2004,ITRS:2003}, is therefore eliminated in the spin transistor.

\begin{figure}
\includegraphics[width=\columnwidth]{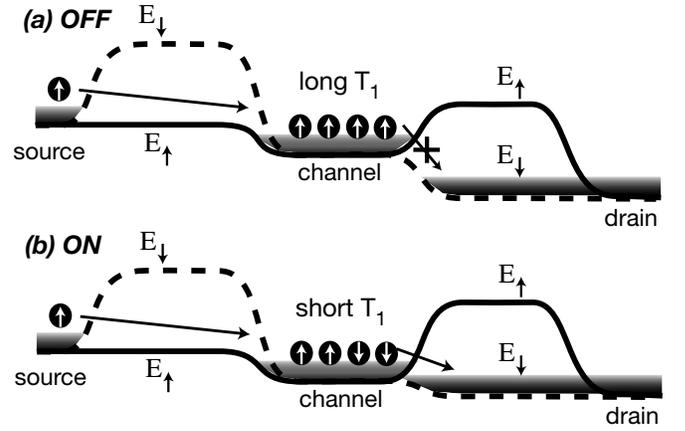}
\caption{(a) Spin transistor in off configuration. (b) Spin transistor in on configuration.}
\label{spintransistor}
\end{figure}

Such a barrier can generate gain when it is used in a transistor geometry\cite{Hall:2003}, as shown in Fig.~\ref{spintransistor}. This spin transistor has very different performance characteristics from MOSFETs. Two oppositely-aligned spin-selective barriers are placed in series, so without any spin-flip in the channel, Fig.~\ref{spintransistor}(a), no source-drain current flows. No significant leakage current comes from tunneling through the barriers of the spin transistor, so the leakage current in the off state originates principally from spin-flip processes. These can occur in the barrier between the source and the channel, in the channel itself, or in the barrier between the channel and the drain. As the channel is the largest region we would expect channel spin relaxation processes to dominate the leakage current when the device is off.  When spin-flip in the channel is rapid, Fig.~\ref{spintransistor}(b), source-drain current flows. For a quantum well channel the current from spin relaxation processes
\begin{equation}
I_{SD} = \frac{A e n }{2\tau_{transit}}\left(1-{\rm e}^{-\tau_{transit}/T_1}\right)\label{current}
\end{equation}
where $I_{SD}$ is the source-drain current, $A$ is the area of the channel, $e$ is the electron charge, $n$ is the two-dimensional electron density in the channel, $\tau_{transit}$ is the carrier transit time through the device, and $T_1$ is the longitudinal relaxation time of the magnetization (half the spin-flip time for individual carriers).   In the limit $T_1\gg \tau_{transit}$ this becomes $I_{SD} = Aen/2T_1$ and the on-off ratio for the transistor is the ratio of $T_1$ in the off state to $T_1$ in the on state (independent of the gate capacitance).

Spin relaxation in the quantum well is controlled by the gate electric field $E$, and the spin relaxation rate is proportional to $E^2$. In the absence of an applied electric field the spin relaxation rate in (110) zincblende quantum wells is quite long\cite{Ohno:1999,Karimov:2003}. The dominant relaxation mechanism in zincblende quantum wells, precessional decoherence, does not contribute, leaving residual spin relaxation from stray electric fields, from spin-flip scattering processes, and from nuclear interactions. Although the limits of these mechanisms are not well-known, spin lifetimes in excess of 100~ns have been observed in GaAs, and we take a spin lifetime of 1~$\mu$s, corresponding to stray electric fields of 200~V/cm (or drift velocities of $\sim 10^7$~cm/s) in the structure of Ref.~\onlinecite{Hall:2003}. The lower limit of the spin lifetime achievable by electric-field tuning is also not known, although tuned times shorter than 10~ps have been achieved\cite{Hall:2005}. The spin lifetime desired for the on state (here assumed to be 10~ps) determines the electric field in the on state, $E_{on}$. The threshold voltage is then 
\begin{equation}
V_{th} = E_{on}D,
\end{equation}
where $D$ is the thickness of the channel quantum well.

In CMOS FETs to keep the leakage current low either a high barrier  (larger $E_{on}$) or thick barrier (larger gate length, and thus larger $A$) is required. Thus $E_{on}$ depends indirectly on $A$. The gate capacitance $C_g$ depends on the channel area and a thickness $d$, determined by the oxide layer, 
\begin{equation}
C_g = \epsilon_0 \epsilon_{r}A/d,
\end{equation}
where $\epsilon_0$ is the permittivity of vacuum and $\epsilon_{r}$ the relative dielectric constant of the region of gate voltage drop. Thus in CMOS FETs $V_{th}$ depends indirectly on $C_g$. No such connection between $V_{th}$ and $C_g$ is apparent for the spin transistor. Furthermore, for the spin transistor, the thickness $d$ is the quantum well thickness $D$.

Gate switching speeds are determined by the time required to charge the capacitor on the next transistor (intrinsic switching delay), hence $\tau_{switch}=V_{th}C_g/(I_{SD,sat})$. For the spin transistor both $C_g$ and $I_{SD,sat}$ are proportional to the channel area, and $V_{th}$ is independent of it, therefore the gate switching time for a fixed on-off ratio is {\it independent of the channel area} and depends on
\begin{equation}
\tau_{switch} = 2 E_{on} T_{1,on} \epsilon_0\epsilon_{sc}/en.\label{gatetime}
\end{equation} 
The power-delay product for a fixed on-off ratio, $ C_g V_{th}^2$,
shrinks proportionally as the area shrinks, as does the leakage current in the off state. Independent of specific designs these scaling features can be summarized as a switching speed and on-off ratio independent of device area, and a power dissipation from both dynamic sources (switching energy) and static sources (leakage current) that is proportional to device area. These very different scaling relations from MOSFETs imply that the performance of spin-based transistors will improve as they become smaller.

Although the scaling relationships indicate that a sufficiently small spin transistor can be superior to a MOSFET, a comparison with a specific design (such as that of Ref.~\onlinecite{Hall:2003}) provides a current benchmark. In Ref.~\onlinecite{Hall:2003} a doping level of $n=2\times 10^{11}$~cm$^{-2}$ in the channel was chosen, but a factor of ten larger doping still permits the spin filtering into and out of the channel to be efficient. The comparison here will use $n=2\times 10^{12}$~cm$^{-2}$.  
An applied electric field of 50~kV/cm across an InAs/AlSb quantum well that is ~200\AA\ thick reduces the $T_1$ to 10~ps, corresponding to $V_{th}=100$~mV, compared with a projected value of $400$~mV for LSTP CMOS in 2018. The lower threshold voltage for the spin transistor is an indication of the small energies required to relax spins. A 1~meV spin splitting from a magnetic field can cause a spin to completely reorient by precession in only 1~ps. A threshold voltage as large as 100~mV is needed only because spin relaxation occurs from internal magnetic fields generated indirectly by the gate electric field through the spin-orbit interaction.

To evaluate the dynamic power dissipation (determined by the power-delay product) a channel area must be chosen. For a gate length of 10~nm and width of 1~$\mu$m the $C_g = 5\times 10^{-17}$F (5 times lower than a 2018 LSTP CMOS\cite{ITRS:2003} transistor of the same gate length and width) and the power-delay product is $ 5\times 10^{-19}$J, compared to the 500 times larger value for a 2018 LSTP CMOS transistor.

\begin{figure}
\includegraphics[width=\columnwidth]{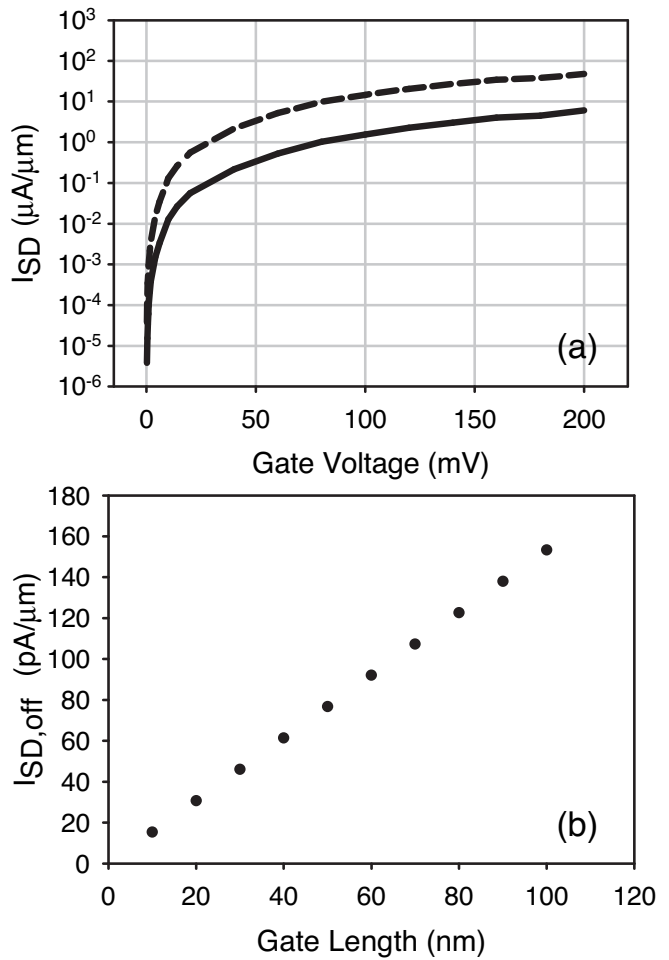}
\caption{(a) Current-voltage relationship for spin transistors with channel lengths of (dashed line) 100~nm and (solid line) 10~nm. (b) Leakage current as a function of channel length. }
\label{sd-leakage}
\end{figure}

Figure~\ref{sd-leakage} shows current-voltage curves for spin transistors with differing channel lengths and reflects the scaling behavior of the static power dissipation.   Fig.~\ref{sd-leakage}(a) shows that, as the channel length is reduced from $100$~nm to $10$~nm, the current in the off state is reduced correspondingly. Fig.~\ref{sd-leakage}(b) shows the dependence of the source-drain leakage current on the channel length, indicating that as the channel is made shorter the leakage {\it decreases}. CMOS FETs, in contrast, have increasing source-drain leakage currents as the channel length is decreased. Compared with 2018 CMOS, with $100$pA/$\mu$m leakage currents, the spin transistor will have a lower leakage current for channel lengths smaller than 60~nm. For the 10~nm long structure described above the leakage current is 6 times smaller, leading to 6 times less static power dissipation.

The above quantities predict a switching time $\tau_{switch}$, from Eq.~\ref{gatetime}, of 3~ps, independent of the channel length or width. This switching time is longer than the 2018 LSTP CMOS value of 0.3~ps. A summary of the compared quantities in Table~\ref{sumtab} indicates that the spin transistor compares favorably with 2018 LSTP CMOS for all properties except the switching time.  One strategy for reducing the switching time would be to increase the threshold voltage (which, however, also increases the gate switching energy). A better approach may be to use a material with a larger spin-orbit interaction strength (such as InSb or an InAs/GaSb superlattice).

\begin{table}
\caption[]{Summary of the comparison between the spin transistor design of Ref.~\onlinecite{Hall:2003} and 2018 LSTP CMOS\cite{ITRS:2003}.}
\begin{tabular} {lcc}
&spin&CMOS\\
\hline
gate length (nm)&10&10\\
gate capacitance $C_g$ (fF/$\mu$m)&0.05&0.25\\
threshold voltage $V_{th}$ (V)&0.1&0.4\\
static leakage current $I_{sd,leak}$ (pA/$\mu$m)&16&100\\
power-delay product (eV/$\mu$m)&3&1500\\
switching time $\tau_{switch}$ (ps)&3&0.3\\
\end{tabular}
\label{sumtab}
\end{table}

An in-depth comparison of a spin transistor design with CMOS design goals for 2018 indicates that, due to their reliance on spin-based switching, the spin transistors can be expected to have superior dynamic and static power dissipation properties. Switching times in a particular spin transistor design (Ref.~\onlinecite{Hall:2003}) are longer than those of 2018 CMOS, but can be brought more in line by increasing the channel doping. The superior switching time of 2018 CMOS  is predicated on the ability to achieve 10$^7$ on-off ratios in devices with the above characteristics, whereas the estimated spin transistor on-off ratio is 10$^5$. Increasing the on-off ratio to 10$^7$ in spin transistors by lengthening the off spin lifetimes would require room-temperature spin lifetimes $\sim 100\mu$s. Although spin lifetimes in excess of 1~ms have been measured in quantum dots at low temperature\cite{Abstreiter:2004}, achieving such lifetimes at room temperature may be very challenging. Our results rely on the development of suitable spin-dependent barrier contacts\cite{HM,Voskoboynikov:2000,deAndradaeSilva:1999,Koga:PRL2002,Ting:2002,Hall:2003}. The 2018 semiconductor roadmap numbers, however, all correspond to goals with no known solution at the present time. 

We acknowledge stimulating conversations with T. F. Boggess. This work was supported by DARPA/ARO DAAD19-01-1-0490, DARPA MDA972-01-C-0002, the National Science Foundation through Grant No. ECS 03-22021, and the Natural Sciences and Engineering Research Council of Canada.

\end{document}